\documentclass[12pt,preprint]{aastex}
\usepackage{graphicx}
\usepackage{natbib}
\shorttitle{Constraining properties of rapidly rotating neutron
stars} \shortauthors{Krastev et al.}

\begin{document}
\title{Constraining properties of rapidly rotating neutron stars using data from heavy-ion collisions}
\author{Plamen G. Krastev, Bao-An Li, and Aaron Worley} \affil{Department of Physics,
Texas A\&M University-Commerce, Commerce, TX 75429, U.S.A.}
\email{Plamen\_Krastev@tamu-commerce.edu,
Bao-An\_Li@tamu-commerce.edu, aworley@leo.tamu-commerce.edu}
%\author{Bao-An Li} \affil{Texas A\&M University-Commerce} \email{Bao-An\_Li@tamu-commerce.edu}
%\author{Aaron Worley} \affil{Texas A\&M University-Commerce} \email{aworley@leo.tamu-commerce.edu}

\begin{abstract}
Properties, structure, and thermal evolution of neutron stars are
determined by the equation of state of stellar matter. Recent data
on isospin-diffusion and isoscaling in heavy-ion collisions at
intermediate energies as well as the size of neutron skin in
$^{208}Pb$ have constrained considerably the density dependence of
the nuclear symmetry energy and, in turn, the equation of state of
neutron-rich nucleonic matter. These constraints could provide
useful information about the global properties of rapidly rotating
neutron stars. Models of rapidly rotating neutron stars are
constructed applying several nucleonic equations of state.
Particular emphasis is placed on configurations rotating rigidly at
$716$ and $1122Hz$. The range of allowed hydrostatic equilibrium
solutions is determined and tested for stability. The effect of
rotation on the internal composition and thermal properties of
neutron stars is also examined. At a given rotational frequency,
each equation of state yields a range of possible neutron stars
configurations restricted by the Keplerian (mass-shedding) limit,
corresponding to the maximal circumferential radius, and the limit
due to the onset of instabilities with respect to axial-symmetric
perturbations, corresponding to the minimal equatorial radius of a
stable neutron star model. We show that the mass of a neutron star
rotating uniformly at $1122Hz$ is between $1.7$ and $2.1M_{\sun}$.
Central stellar density and proton fraction decrease with increasing
rotational frequency with respect to static models, and depending on
the exact stellar mass and angular velocity, can drop below the
Direct Urca threshold thus closing the fast cooling channel.
\end{abstract}

\maketitle

\section{Introduction}

Neutron stars are one of the most exotic objects in the universe.
Matter in their cores is compressed to huge densities ranging from
the density of normal nuclear matter, $\rho_0\approx{0.16}fm^{-3}$,
to an order of magnitude higher~\citep{Glendenning:2000a}. The
number of baryons forming a neutron star is in the order of
$A\approx{10^{57}}$. Understanding properties of matter under such
extreme conditions of density (and pressure) is still far from
complete and represents one of the most important but also
challenging problems in modern physics.

Strictly speaking, neutron stars are associated with two classes of
astrophysical objects~\citep{Weber:1999a}. Pulsars belong to the
first class and are generally accepted to be rotating neutron stars.
X-ray sources belong to the second class and some of them are
neutron stars on close binary orbit with an ordinary star (e.g., Her
X-1 and Vela X-1)~\citep{Weber:1999a}. Because of their strong
gravitational binding neutron stars can rotate very
fast~\citep{Bejger:2006hn}. The first millisecond pulsar
PSR1937+214, spinning at $\nu=641 Hz$~\citep{Backer:1982}, was
discovered in 1982, and during the next decade or so almost every
year a new one was reported. In the recent years the situation
changed considerably with the discovery of an anomalously large
population of millisecond pulsars in globular
clusters~\citep{Weber:1999a}, where the density of stars is roughly
1000 times that in the field of the galaxy and which are therefore
very favorable sites for formation of rapidly rotating neutron stars
which have been spun up by the means of mass accretion from a binary
companion. Presently the number of the observed pulsars is close to
2000, and the detection rate is rather high.

In 2006~\citet{Hessels:2006ze} reported the discovery of a very
rapid pulsar J1748-2446ad, rotating at $\nu=716Hz$ and thus breaking
the previous record (of $641Hz$). However, even this high rotational
frequency is too low to affect the structure of neutron stars with
masses above $1M_{\sun}$~\citep{Bejger:2006hn}. Such pulsars belong
to the slow-rotation regime since their frequencies are considerably
lower than the Kepler (mass-shedding) frequency $\nu_k$. (The
mass-shedding, or Kepler, frequency is the highest possible
frequency for a star before it starts to shed mass at the equator.)
Neutron stars with masses above $1M_{\sun}$ enter the rapid-rotation
regime if their rotational frequencies are higher than $1000Hz$
\citep{Bejger:2006hn}. A recent report by~\citet{Kaaret:2006gr}
suggests that the X-ray transient XTE J1739-285 contains the most
rapid pulsar ever detected rotating at $\nu=1122Hz$. This discovery
has reawaken the interest in building models of rapidly rotating
neutron stars~\citep{Bejger:2006hn}.

Since neutron stars are objects of extremely condensed matter, the
geometry of space-time is considerably altered from that of a flat
space. Therefore, the construction of realistic models of neutron
stars has to be done in the framework of General
Relativity~\citep{Weber:1999a,Glendenning:2000a}. Detailed
knowledge of the equation of state (EOS) of stellar matter over a
very wide range of densities is required for solving the
neutron-star structure equations. At present time the behavior of
matter under extreme densities such as those found in the
interiors of neutron stars is still highly uncertain and relies
upon, often, rather controversial theoretical predictions.
Fortunately, heavy-ion reactions provide a unique means to
constrain the EOS of dense nuclear matter in terrestrial
laboratories~\citep{Danielewicz:2002pu}. In particular, one of the
main sources of uncertainty in the stellar matter EOS is the
density dependence of the nuclear symmetry
energy~\citep{Lattimer:2004pg,Steiner:2004fi,Krastev:2006ii},
$e_{sym}(\rho)$, which is the difference between the energies of
pure neutron and symmetric nuclear matter. Due to its importance
for the neutron star structure determining the density dependence
of the nuclear symmetry energy has been a priority goal for the
intermediate energy heavy-ion community. Although extracting the
symmetry energy is not an easy task due to the complicated role of
the isospin degree of freedom in reaction dynamics, several
promising probes of the symmetry energy have been
suggested~\citep{Li:1997rc,Li:2000bj,Li:2002qx,Li:1997px} (see
also~\citep{Li:2001a,Danielewicz:2002pu,Baran:2004ih} for
reviews). Some significant progress have been made recently in
determining the density dependence of $e_{sym}$ at subsaturation
densities using: (1) isospin diffusion~\citep{Tsang:2004} and
isoscaling~\citep{Tsang:2001,Shetty:2007} in heavy-ion reactions
at intermediate
energies~\citep{Shi:2003np,Chen:2005a,Steiner:2005rd,Li:2005jy},
and (2) sizes of neutron skins in heavy
nuclei~\citep{Steiner:2004fi,Horowitz:2000xj,Horowitz:2002mb,ToddRutel:2005fa}.
At supranormal densities, a number of potential probes of the
symmetry energy have been proposed although there is not much data
available yet~\citep{Chen:2007}.

While global properties of spherically symmetric static
(non-rotating) neutron stars have been studied extensively
\citep{Lattimer:2000kb,Lattimer:2004pg,Prakash:2001rx,Yakovlev:2004iq,Heiselberg:2000dn,Heiselberg:1999mq,Steiner:2004fi,Krastev:2006ii},
properties of (rapidly) rotating neutron stars have been
investigated to lesser extent. Models of (rapidly) rotating
neutron stars have been constructed only by several research
groups with various degree of approximation
\citep{Hartle:1967he,Hartle:1968si,Friedman:1986tx,Bombaci:2000rc,1990ApJ...355..241L,1989MNRAS.237..355K,1994ApJ...424..823C,Stergioulas:1994ea,Stergioulas:1997ja,1993A&A...278..421B,1998PhRvD..58j4020B,Weber:1999a,2002A&A...381L..49A}
(see \citet{Stergioulas:2003yp} for a review). In this paper we
combine recently obtained data on isospin diffusion, information
from flow observables, studies of neutron skin data of $^{208}Pb$
and other information to constrain global properties of rotating
neutron stars. Applying several nucleonic equations of state and
the $RNS$\footnote{Thanks to Nikolaos Stergioulas the $RNS$ code
is available as a public domain program at
http://www.gravity.phys.uwm.edu/rns/} code developed and made
available to the public by Nikolaos
Stergioulas~\citep{Stergioulas:1994ea}, we construct one-parameter
2-D stationary configurations of rapidly rotating neutron stars.
The computation solves the hydrostatic and Einstein field
equations for mass distributions rotating rigidly under the
assumptions of stationary and axial symmetry about the rotational
axis, and reflection symmetry about the equatorial plane. This
work is organized in the following way: after the introduction in
Section 1, we review briefly the structure equations of both
static and rotating neutron stars on Section 2. Our numerical
results are presented and discussed in Section 3. We conclude with
a short summary in Section 4.

\section{Equations of neutron star structure}

In what follows we present a brief review of structure equations of
both static and rotating neutron stars. As already mentioned in the
introduction neutron stars are objects of extremely compressed
matter and therefore proper understanding of their properties
requires application of both General Relativity and the theories of
dense matter, which constitute nuclear and particle physics problem.
In this respect neutron stars provide a direct link between two of
the frontiers of modern physics - General Relativity and strong
interactions in dense matter~\citep{Weber:1999a}. The connection
between both branches of physics is provided by Einstein's field
equations
\begin{equation}\label{eq.1}
G^{\mu\nu}=R^{\mu\nu}-\frac{1}{2}g^{\mu\nu}R=8\pi
T^{\mu\nu}(\epsilon,P(\epsilon)),
\end{equation}
$(\mu,\nu=0,1,2,3)$ which couple the Einstein curvature tensor,
$G^{\mu\nu}$, to the energy-momentum tensor,
\begin{equation}\label{eq.2}
T^{\mu\nu} = (\epsilon+P)u^{\mu}u^{\nu}+Pg^{\mu\nu},
\end{equation}
of stellar matter. In the above equations $P$ and $\epsilon$ denote
pressure and mass energy density, while $R^{\mu\nu}$, $g^{\mu\nu}$,
and $R$ denote the Ricci tensor, the metric tensor, and the Ricci
scalar curvature respectively~\citep[see
e.g.,][]{Glendenning:2000a}. In Eq.~(\ref{eq.2}) $u^\mu$ is the unit
time-like four-velocity satisfying $u^{\mu}u_{\mu}=-1$. The tensor
$T^{\mu\nu}$ contains the EOS of stellar matter in the form
$P(\epsilon)$. In general, Einstein's field equations and those of
the nuclear many-body problem were to be solved simultaneously since
the baryons and quarks follow the geodesics of the curved space-time
whose geometry, determined by the Einstein's field equations, is
coupled to the total mass energy density of
matter~\citep{Weber:1999a}. In the case of neutron stars, as for all
astrophysical situations for which the long-range gravitational
force can be separated from the short-range strong force, the
deviation from flat space-time at the length-scale of the strong
interactions ($\sim 1fm$) is practically zero up to the highest
densities achieved in the neutron star interiors. (This is not to be
confused with the global length-scale of neutron stars ($\sim 10km$)
for which $M/R\sim 0.3$ depending on the star's mass (in units
$c=G=1$ so that $M_{\sun}\approx 1.475km$).) In other words, gravity
curves space-time only on a macroscopic scale but to a very good
approximation leaves it flat on a microscopic scale. To achieve an
appreciable curvature on a microscopic level at which the strong
interactions dominate the particle dynamics mass densities greater
than $\sim 10^{40} g\hspace{1mm}cm^{-3}$ would be
necessary~\citep{Weber:1999a,Thorne1966a}. Under this circumstances
the problem of constructing models of neutron stars separates into
two distinct tasks. First, the short-range effects of the nuclear
forces are described by the principles of many-body nuclear physics
in a local inertial frame (co-moving proper reference frame) in
which space-time is flat. Second, the coupling between the
long-range gravitational force and matter is accounted for by
solving the general relativistic equations for the gravitational
field described by the curvature of space-time, leading to the
global structure of stellar configurations.

\subsection{Static stars}

In the case of spherically symmetric static (non-rotating) stars the
metric has the famous Schwarzschild form:
\begin{equation}\label{eq.3}
ds^2=-e^{2\phi(r)}dt^2+e^{2\Lambda(r)}dr^2+r^2(d\theta^2+\sin^2\theta
d\phi^2),
\end{equation}
$(c=G=1)$ where the metric functions $\phi(r)$ and $\Lambda(r)$ are
given by:
\begin{equation}\label{eq.4}
e^{2\Lambda(r)}=(1-\gamma(r))^{-1},
\end{equation}
\begin{equation}\label{eq.5}
e^{2\phi(r)}=e^{-2\Lambda(r)}=(1-\gamma(r))\quad r>R_{star},
\end{equation}
with
\begin{equation}\label{eq.6}
\gamma(r)=\left\{
\begin{array}{l l}
\frac{2m(r)}{r} & \quad \mbox{if $r<R_{star}$}\\\\
\frac{2M}{r} & \quad \mbox{if $r>R_{star}$}
\end{array}
\right.
\end{equation}
For a static star Einstein's field equations (Eq.~(\ref{eq.1}))
reduce then to the familiar Tolman-Oppenheimer-Volkoff equation
(TOV)~\citep{Tolman:1939jz,PhysRev.55.374}:
\begin{equation}\label{eq.7}
\frac{dP(r)}{dr}=-\frac{\epsilon(r)m(r)}{r^2}
\left[1+\frac{P(r)}{\epsilon(r)}\right]
\left[1+\frac{4\pi{r^3}p(r)}{m(r)}\right]
\left[1-\frac{2m(r)}{r}\right]^{-1}
\end{equation}
where the gravitational mass within a sphere of radius $r$ is
determined by
\begin{equation}\label{eq.8}
\frac{dm(r)}{dr}=4\pi\epsilon(r)r^{2}dr
\end{equation}
The metric function $\phi(r)$ is determined through the following
differential equation:
\begin{equation}\label{eq.9}
\frac{d\phi(r)}{dr}=-\frac{1}{\epsilon(r)+P(r)}\frac{dP(r)}{dr},
\end{equation}
with the boundary condition at $r=R$
\begin{equation}\label{eq.10}
\phi(r=R)=\frac{1}{2}\ln\left(1-\frac{2M}{R}\right)
\end{equation}

To proceed to the solution of these equations, it is necessary to
provide the EOS of stellar matter in the form $P(\epsilon)$.
Starting from some central energy density $\epsilon_c=\epsilon(0)$
at the center of the star $(r=0)$, and with the initial condition
$m(0)=0$, the above equations can be integrated outward until the
pressure vanishes, signifying that the stellar edge is reached.
Some care should be taken at $r=0$ since, as seen above, the TOV
equation is singular there. The point $r=R$  where the pressure
vanishes defines the radius of the star and
$M=m(R)=4\pi\int_0^R\epsilon(r')r'^2dr'$ its gravitational mass.

For a given EOS, there is a unique relationship between the stellar
mass and the central density $\epsilon_c$. Thus, for a particular
EOS, there is a unique sequence of stars parameterized by the
central density (or equivalently the central pressure $P(0)$).

\subsection{Rotating stars}

Equations of stellar structure of (rapidly) rotating neutron stars
are considerably more complex than those of spherically symmetric
stars~\citep{Weber:1999a}. These complications arise due to the
rotational deformations in rotating stars (i.e., flattening at the
poles and bulging at the equator), which lead to a dependence of the
star's metric on the polar coordinate $\theta$. In addition,
rotation stabilizes the star against gravitational collapse and
therefore rotating neuron stars are more massive than static ones. A
larger mass, however, causes greater curvature of space-time. This
renders the metric functions frequency-dependent. Finally, the
general relativistic effect of dragging the local inertial frames
implies the occurrence of an additional non-diagonal term,
$g^{t\phi}$, in the metric tensor $g^{\mu\nu}$. This term imposes a
self-consistency condition on the stellar structure equations, since
the degree at which the local inertial frames are dragged along by
the star, is determined by the initially unknown stellar properties
like mass and rotational frequency~\citep{Weber:1999a}.

Here we outline briefly the equations solved by the $RNS$ code. The
coordinates of the stationary, axial symmetric space-time used to
model a (rapidly) rotating neutron star are defined through a
generalization of Bardeen's metric~\citep{Stergioulas:2003yp}:
\begin{eqnarray}\label{eq.11}
ds^2&=&-e^{\gamma+\rho}dt^2+e^{2\alpha}(dr^2+r^2d\theta^2)\nonumber\\
&+&e^{\gamma-\rho}r^2\sin^2\theta(d\phi-\omega dt^2),
\end{eqnarray}
where the metric potentials $\gamma$, $\rho$, $\alpha$, and the
angular velocity of the stellar fluid relative to the local inertial
frame, $\omega$, are functions of the quasi-isotropic radial
coordinates, $r$, and the polar angle $\theta$ only. The matter
inside a rigidly rotating star is approximated as a perfect
fluid~\citep{Stergioulas:2003yp}, whose energy momentum tensor is
given by Eq.~(\ref{eq.2}). The proper velocity of matter,
$\upsilon$, relative to the local Zero Angular Momentum Observer
(ZAMO)~\citep{2002A&A...382..939O} is defined as
\begin{equation}\label{eq.12}
\upsilon=r\sin(\theta)(\Omega-\omega)e^{-\rho(r)}
\end{equation}
with $\Omega=u^3/u^0$ the angular velocity of a fluid element. The
four-velocity is given by
\begin{equation}\label{eq.13}
u^{\mu}=\frac{e^{-(\gamma+\rho)/2}}{\sqrt{(1-\upsilon^2)}}(1,0,0,\Omega)
\end{equation}
In the above equation the function $(\gamma+\rho)/2$ represents the
relativistic generalization of the Newtonian gravitational
potential, while $\exp[(\gamma+\rho)/2]$ is a time dilation factor
between an observer moving with angular velocity $\omega$ and one at
infinity. Substitution of Eq.~(\ref{eq.13}) into Einstein's fields
equations projected onto the ZAMO reference frame gives three
elliptic partial differential equations for the metric potentials
$\gamma$, $\rho$, and $\omega$, and two linear ordinary differential
equations for the metric potential $\alpha$. Technically, the
elliptic differential equations for the metric functions are
converted into integral equations which are then solved iteratively
applying Green's function
approach~\citep{1989MNRAS.237..355K,Stergioulas:2003yp}.

From the relativistic equations of motion, the equations of
hydrostatic equilibrium for a barotropic fluid may be obtained
as~\citep{Stergioulas:2003yp,2002A&A...382..939O}:
\begin{equation}\label{eq.14}
h(P)-h_p=\frac{1}{2}[\omega_p+\rho_p-\gamma-\rho-\ln(1-\upsilon^2)+A^2(\omega-\Omega_c)^2],
\end{equation}
with $h(P)$ the specific enthalpy, $P_p$ the re-scaled pressure,
$h_p$ the specific enthalpy at the pole, $\gamma_p$ and $\rho_p$ the
values of the metric potentials at the pole, $\Omega_c=r_e\Omega$,
and $A$ a rotational constant~\citep{Stergioulas:2003yp}. The
subscripts $p$, $e$, and $c$ label the corresponding quantities at
the pole, equator and center respectively. The $RNS$ code solves
iteratively the integral equations for $\rho$, $\gamma$ and
$\omega$, and the ordinary differential equation for the metric
function $\alpha$ coupled with Eq.~(\ref{eq.14}) and the equations
for hydrostatic equilibrium at the stellar center and equator (given
$h(P_c)$ and $h(P_e)=0$) to obtain $\rho$, $\gamma$, $\alpha$,
$\omega$, the equatorial coordinate radius $r_e$, angular velocity
$\Omega$, energy density $\epsilon$, and pressure $P$ throughout the
star.

\section{Results and discussion}
We compute properties of (rapidly) rotating neutron stars employing
several nucleonic EOSs and the $RNS$ code. The EOSs applied here are
shown in Fig.~1.
\begin{figure}[!t]
\centering
\includegraphics[totalheight=3.6in]{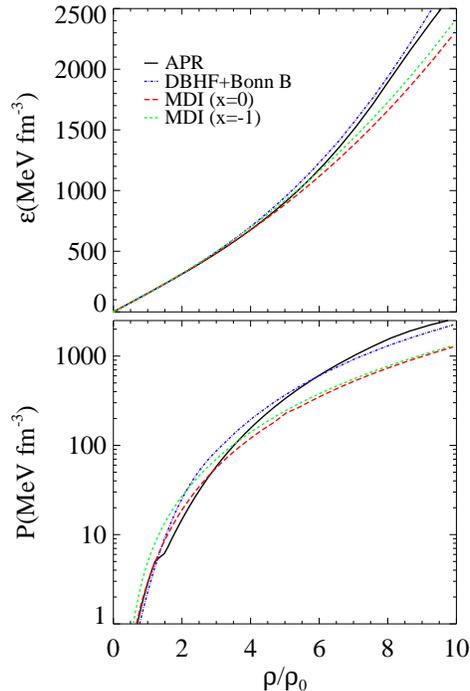}
\vspace{5mm} \caption{(Color online) Equation of state. The upper
frame shows the mass energy density as a function of baryon density
(in units of $\rho_0$) and the lower frame shows the total pressure
(including the lepton contributions) versus baryon density. (The
``dip'' exhibited by the density curve of the APR EOS is due to a
phase transition from low density phase (LDP) to high density phase
(HDP). See \citep{Akmal:1998cf} for details.)}
\end{figure}
We pay particular attention to the EOS computed with the MDI
interaction~\citep{Das:2002fr} since its symmetry energy is
constrained in the sub-saturation density region by the available
nuclear laboratory data. The EOS of symmetric nuclear matter for the
MDI interaction is constrained up to five times the normal nuclear
matter density by the available data on collective flow in
relativistic heavy-ion collisions. The parameter $x$ is introduced
in the single-particle potential of the MDI EOS to reflect the
largely uncertain density dependence of the nuclear symmetry energy
$e_{sym}(\rho)$ as predicted by various many-body approaches. Since,
as demonstrated by \citet{Li:2005jy} and \citet{Li:2005sr}, only
equations of state with $x$ between -1 and 0 have symmetry energies
consistent with the isospin diffusion data and measurements of the
skin thickness of $^{208}Pb$, we thus consider only these two
limiting cases as boundaries of the possible rotating neutron star
models. Moreover, it is interesting to note that the symmetry energy
extracted very recently from the isoscaling analyses of heavy-ion
reactions is consistent with the MDI calculation using $x=0$
\citep{Shetty:2007}. The MDI EOS has been recently applied to
constrain the neutron-star radius~\citep{Li:2005sr} with a suggested
range compatible with the best estimates from observations. In
addition, it has been also used to constrain a possible time
variation of the gravitational constant $G$~\citep{Krastev:2007en}
via the gravitochemical heating formalism developed by
\citet{Jofre:2006ug}. In Fig.~1 we also show results by
\citet{Akmal:1998cf} with the $A18+\delta\upsilon+UIX*$ interaction
(APR) and recent Dirac-Brueckner-Hartree-Fock (DBHF)
calculations~\citep{Alonso:2003aq,Krastev:2006ii} with Bonn B
One-Boson-Exchange (OBE) potential (DBHF+Bonn
B)~\citep{Machleidt:1987hj}. Below the baryon density of
approximately $0.07fm^{-3}$ the equations of state shown in Fig.~1
are supplemented by a crustal EOS, which is more suitable for the
low density regime. Namely, we apply the EOS by~\citet{PRL1995} for
the inner crust and the one by~\citet{HP1994} for the outer crust.
At the highest densities we assume a continuous functional for the
EOSs employed in this work. (See~\citep{Krastev:2006ii} for a
detailed description of the extrapolation procedure for the
DBHF+Bonn B EOS.) The saturation properties of the nuclear equations
of state used in this paper are summarized in Table 1.

\begin{table}[!h]
\caption{Saturation properties of the nuclear EOSs (for symmetric
nuclear matter) shown in Fig.~1.}
\begin{center}
\begin{tabular}{lccccc}
EOS &  $\rho_0(fm^{-3})$ & $E_s(MeV)$ & $\kappa(MeV)$ & $e_{sym}(\rho_0)(MeV)$ & $m^*(\rho_0)/m$\\
\hline\hline
MDI(x=0)    & 0.160 & -16.08 & 211.00 & 31.62 & 0.67\\
MDI(x=-1)   & 0.160 & -16.08 & 211.00 & 31.62 & 0.67\\
APR         & 0.160 & -16.00 & 266.00 & 32.60 & 0.70\\
DBHF+Bonn B & 0.185 & -16.14 & 259.04 & 33.71 & 0.65\\
\hline
\end{tabular}
\end{center}
{\small The first column identifies the equation of state. The
remaining columns exhibit the following quantities at the nuclear
saturation density: saturation (baryon) density;
energy-per-particle; compression modulus; symmetry energy; nucleon
effective mass to {\it average} nucleon mass ratio (with
$m=938.926MeV$ $c^{-2}$).}
\end{table}

\subsection{Keplerian (and static) sequences}

\begin{figure}[!t]
\centering
\includegraphics[totalheight=4.0in]{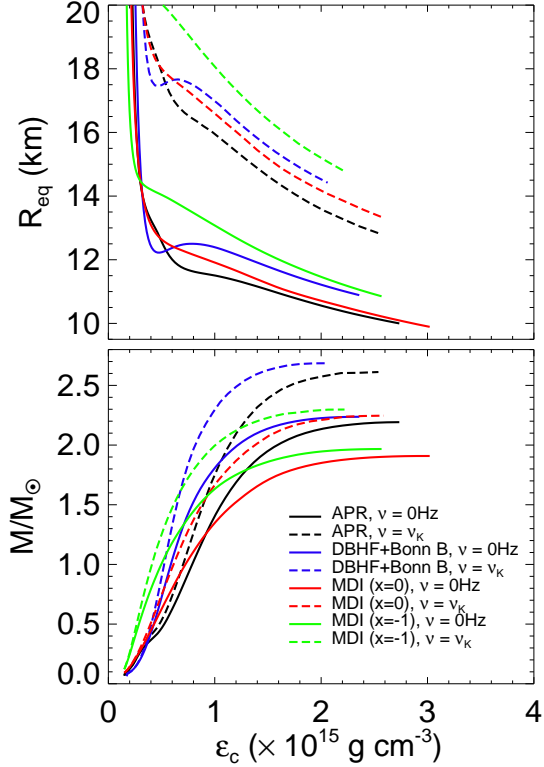}
\vspace{5mm} \caption{(Color online) Neutron star masses and radii.
Neutron star equatorial radii (upper panel) and total gravitational
mass (lower panel) versus central energy density $\epsilon_c$. Both
static (solid lines) and Keplerian (broken lines) models are shown.}
\end{figure}

In what follows we examine the effect of ultra-fast rotation at the
Kepler frequency on the neutron star gravitational masses and radii.
The equilibrium configurations for both static and (rapidly)
rotating neutron stars are parameterized in terms of the central
mass energy density, $\epsilon_c=\epsilon(0)$ (or equivalently
central pressure, $P_c=P(0)$). This functional dependence is shown
in Fig.~2, where we display the stellar {\it equatorial} radius
(upper frame) and total gravitational mass (lower frame) versus
central energy density for the EOSs applied in this work.
Predictions for both static and maximally rotating models are shown.
\begin{figure}[!t]
\centering
\includegraphics[totalheight=2.8in]{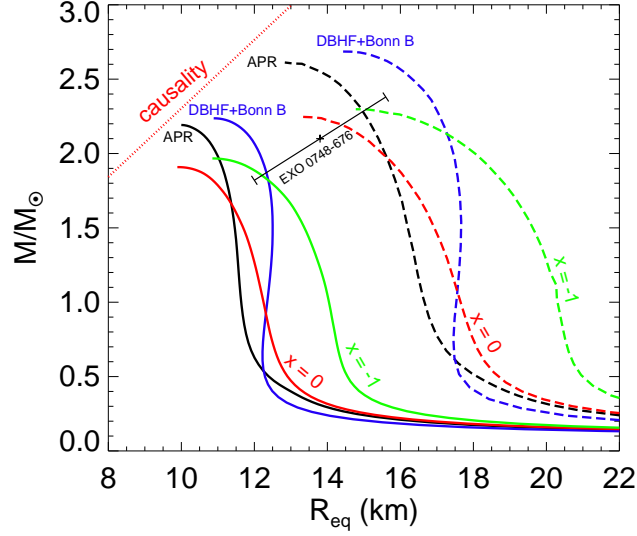}
\caption{(Color online) Mass-radius relation. Both static (solid
lines) and Keplerian (broken lines) sequences are shown. The
$1-\sigma$ error bar corresponds to the measurement of the mass and
radius of EXO 0748-676~\citep{Ozel:2006km}.}
\end{figure}
\begin{figure}[!t]
\centering
\includegraphics[totalheight=2.5in]{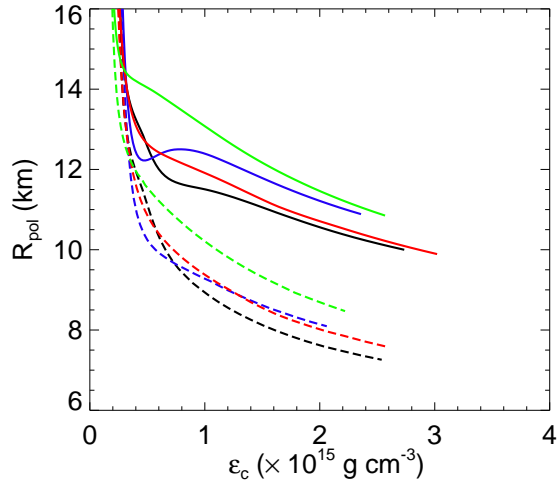}
\caption{(Color online) Neutron-star polar radius versus central
energy density. Both static (solid lines) and Keplerian (broken
lines) sequences are shown. The labeling of the curves is the same
as in Fig.~2.}
\end{figure}
The main feature of the mass-density plot is that there exists a
maximum value of the gravitational mass of a neutron star that a
given EOS can support~\citep[see e.g.,][]{Weber:1999a}. This holds
for both static and (rapidly) rotating stars. The sequences shown in
Fig.~2 terminate at the "maximum mass" point. Comparing the results
for static and rotating stars, it is seen clearly that the rapid
rotation increases noticeably the mass that can be supported against
collapse while lowering the central density of the maximum-mass
configuration. This is what one should expect, since, as already
mentioned, rotation stabilizes the star against the gravitational
pull providing an extra (centrifugal) repulsion. The rotational
effect on the mass-radius relation is illustrated in Fig.~3 where
the gravitational mass is given as a function of the circumferential
radius. For rapid rotation at the Kepler frequency, a mass increase
up to $\sim 17\%$ (Table~3) is obtained, depending on the EOS. The
equatorial radius increases by several kilometers, while the polar
radius decreases by several kilometers (see Fig.~4) leading to an
overall oblate shape of the rotating star. Table 2 summarizes the
properties (masses, radii and central energy densities) of the
maximum-mass nonrotating neutron star configurations. Our studies on
the effect of rapid rotation on the upper mass limits for the four
EOSs considered in the present paper are presented in Table 3. In
each case the upper mass limit is attained for a model at the
mass-shedding limit where $\nu=\nu_k$, with central density $\sim
15\%$ below that of the static model with the largest mass. These
findings are consistent with those by \citet{1984Natur.312..255F}
and \citet{Stergioulas:1994ea}. Table 3 also provides an estimate of
the upper limiting rotation rate of a neutron star. In general,
softer EOSs permit larger rotational frequencies since the resulting
stellar models are more centrally condensed \citep[see
e.g.,][]{1984Natur.312..255F}. In the last column of Table 3 we show
the Kepler frequencies computed via the empirical relation
\begin{table}[!b]
\caption{Maximum-mass static (nonrotating) models.}
\begin{center}
\begin{tabular}{lccc}
EOS &  $M_{max}(M_{\odot})$ & $R(km)$ & $\epsilon_c(\times 10^{15}g\hspace{1mm}cm^{-3})$\\
\hline\hline
MDI(x=0)       & 1.91    &  9.89  & 3.02\\
APR            & 2.19    &  9.98  & 2.73\\
MDI(x=-1)      & 1.97    & 10.85  & 2.57\\
DBHF+Bonn B    & 2.24    & 10.88  & 2.36\\
\hline
\end{tabular}
\end{center}
{\small The first column identifies the equation of state. The
remaining columns exhibit the following quantities for the static
models with maximum gravitational mass: gravitational mass; radius;
central mass energy density.}
\end{table}
\begin{table}[!b]
\caption{Maximum-mass rapidly rotating models at the Kepler
frequency $\nu=\nu_k$.}
\begin{center}
\begin{tabular}{lccccc}
EOS &  $M_{max}(M_{\odot})$ & Increase (\%) & $\epsilon_c(\times
10^{15}g\hspace{1mm}cm^{-3})$ & $\nu_k(Hz)$
& $\nu_k^{FIP}(Hz)$\\
\hline\hline
MDI(x=0)       & 2.25    &  15  & 2.59 & 1742 & 1610\\
APR            & 2.61    &  17  & 2.53 & 1963 & 1699\\
MDI(x=-1)      & 2.30    &  14  & 2.21 & 1512 & 1423\\
DBHF+Bonn B    & 2.69    &  17  & 2.06 & 1685 & 1510\\
\hline
\end{tabular}
\end{center}
{\small The first column identifies the equation of state. The
remaining columns exhibit the following quantities for the maximally
rotating models with maximum gravitational mass: gravitational mass;
its percentage increase over the maximum gravitational mass of
static models; central mass energy density; maximum rotational
frequency; Kepler frequency as computed via the empirical relation
by.}
\end{table}
\begin{equation}\label{eq.15}
\frac{\Omega_k}{10^4s^{-1}}=0.72\left(\frac{M_s}{M_{\odot}}\right)^{1/2}\left(\frac{R_s}{10km}\right)^{-3/2}
\end{equation}
proposed by \citet{Friedman:1989}. The uncertainty of
Eq.~(\ref{eq.15}) is $\sim 10\%$ (see~\cite{HZ1989} for an improved
version of the empirical formula). At the time of constructing the
above relation~\citet{Friedman:1989} did not consider a then unknown
class of minimum period EOSs \citep{Stergioulas:1996} which explains
why the numbers in the last column of Table 3 exhibit larger
deviation from the exact numerical solutions (in column five). This
is particularly pronounced for the APR EOS for which the
approximated Kepler frequency deviates $\sim 14\%$ from the exact
solution. (Note that the APR EOS has the lowest period among the
EOSs considered here.)

\begin{figure}[!h]
\centering
\includegraphics[totalheight=2.8in]{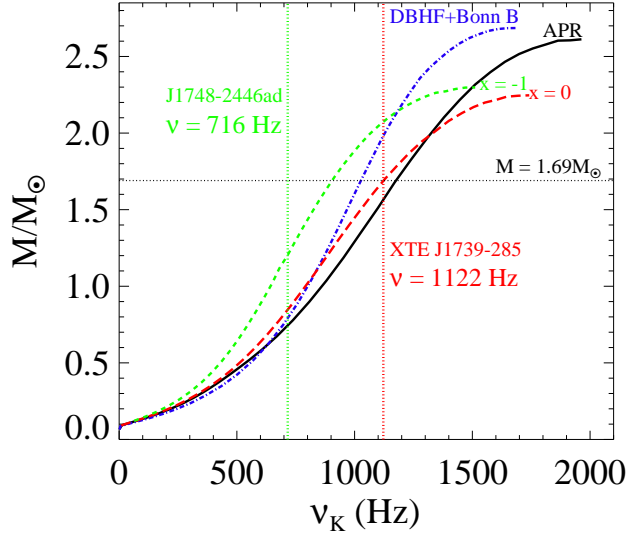}
\caption{(Color online) Mass versus Keplerian (mass-shedding)
frequency $\nu_k$.}
\end{figure}

Fig.~5 displays the neutron star gravitational mass as a function of
the Kepler frequency, $\nu_k$. The vertical doted lines denote the
frequencies of the fastest neutron stars - XTE J1739-285
\citep{Kaaret:2006gr} (red) and J1748-2446ad \citep{Hessels:2006ze}
(green). From this figure one can conclude that only neutron stars
with masses above approximately $1.7M_{\odot}$ can rotate at
$\nu=1122Hz$ (and/or faster). We discuss this further in the next
subsection (3.2).

\subsection{Rotation at 716 and 1122 Hz}

\begin{figure}[!b]
\centering
\includegraphics[totalheight=2.8in]{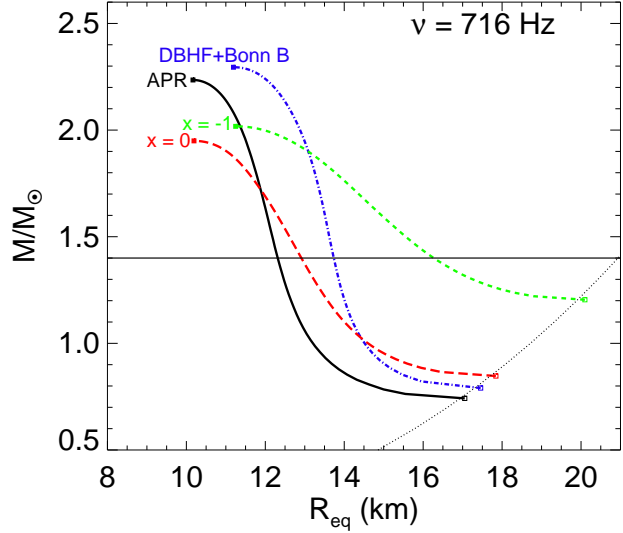}
\caption{(Color online) Gravitational mass versus circumferential
radius for neutron stars rotating at $\nu=716Hz$. See text for
details.}
\end{figure}

\begin{figure}[!t]
\centering
\includegraphics[totalheight=2.8in]{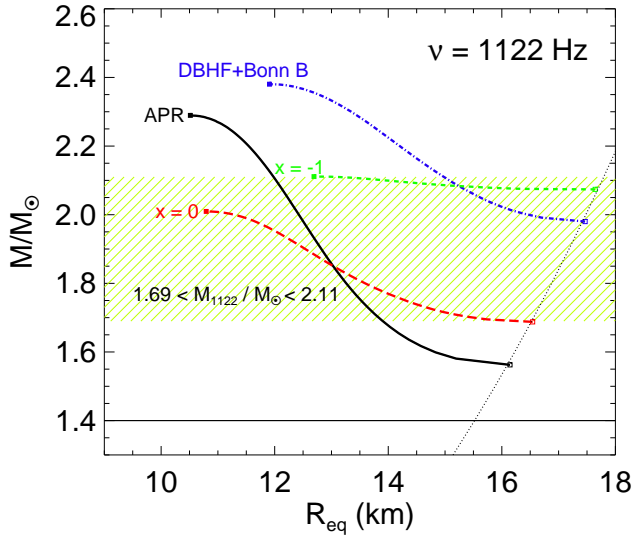}
\caption{(Color online) Gravitational mass versus circumferential
radius for neutron stars rotating at $\nu=1122Hz$.}
\end{figure}

In this subsection we study neutron stars rotating at 716
\citep{Hessels:2006ze} and 1122 Hz \citep{Kaaret:2006gr} which are
the rotational frequencies of the fastest pulsars as of today.
Stability with respect to the mass-shedding from equator implies
that at a given gravitational mass the equatorial radius $R_{eq}$
should be smaller than $R_{eq}^{max}$ corresponding to the Keplerian
limit~\citep{Bejger:2006hn}. The value of $R_{eq}^{max}$ results
from the condition that the frequency of a test particle at circular
equatorial orbit of radius $R_{eq}^{max}$ just above the equator of
the actual rotating star is equal to the rotational frequency of the
star. As reported by~\citep{Bejger:2006hn} the relation between $M$
and $R_{eq}$ at the ``mass-shedding point'' is very well
approximated by the expression for the orbital frequency for a test
particle orbiting at $r=R_{eq}$ in the Schwarzschild space-time
created by a spherical mass. The formula satisfying
$\nu_{orb}^{Schw.}=\nu$, represented by the dotted line in Figs.~6
and 7, is given by
\begin{equation}\label{eq.16}
\frac{1}{2\pi}\left(\frac{GM}{R_{eq}^3}\right)=\nu,
\end{equation}
where $\nu=716Hz$ in Fig.~6 and $\nu=1122Hz$ in Fig.~7 respectively.
This formula for the Schwarzschild metric coincides with the one
obtained in Newtonian gravity for a point mass $M$
\citep{Bejger:2006hn}. Eq.~(\ref{eq.16}) implies
\begin{equation}\label{eq.17}
R_{max}=\chi\left(\frac{M}{1.4M_{\odot}}\right)^{1/3}km,
\end{equation}
with $\chi=20.94$ for rotational frequency $\nu=716Hz$ (Fig.~6) and
$\chi=15.52$ for $\nu=1122Hz$ (Fig.~7).

In Figs.~6 and 7 we observe that the range of the allowed masses
supported by a given EOS for rapidly rotating neutron stars becomes
narrower than the one of static configurations. This effect becomes
stronger with increasing frequency and depends upon the EOS. For
instance, for models rotating at $1122Hz$ (Fig.~7) for the x=-1 EOS
the allowed mass range is $\sim 0.1M_{\sun}$. Since predictions from
the x=0 and x=-1 EOSs represents the limits of the neutron star
models consistent with the nuclear data from terrestrial
laboratories, we conclude that the mass of the neutron star in XTE
J1739-285 is between 1.7 and $2.1M_{\sun}$.

\subsection{Rotation and proton fraction}

\begin{figure}[!b]
\centering
\includegraphics[totalheight=3.6in]{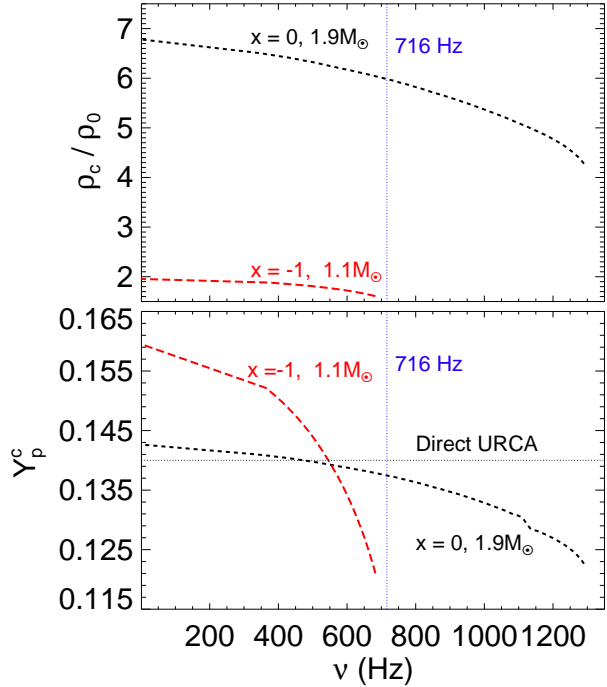}
\vspace{5mm} \caption{(Color online) Density (upper panel) and
proton fraction (lower panel) versus rotational frequency for fixed
neutron star mass.}
\end{figure}

\begin{figure}[!t]
\centering
\includegraphics[totalheight=4.0in]{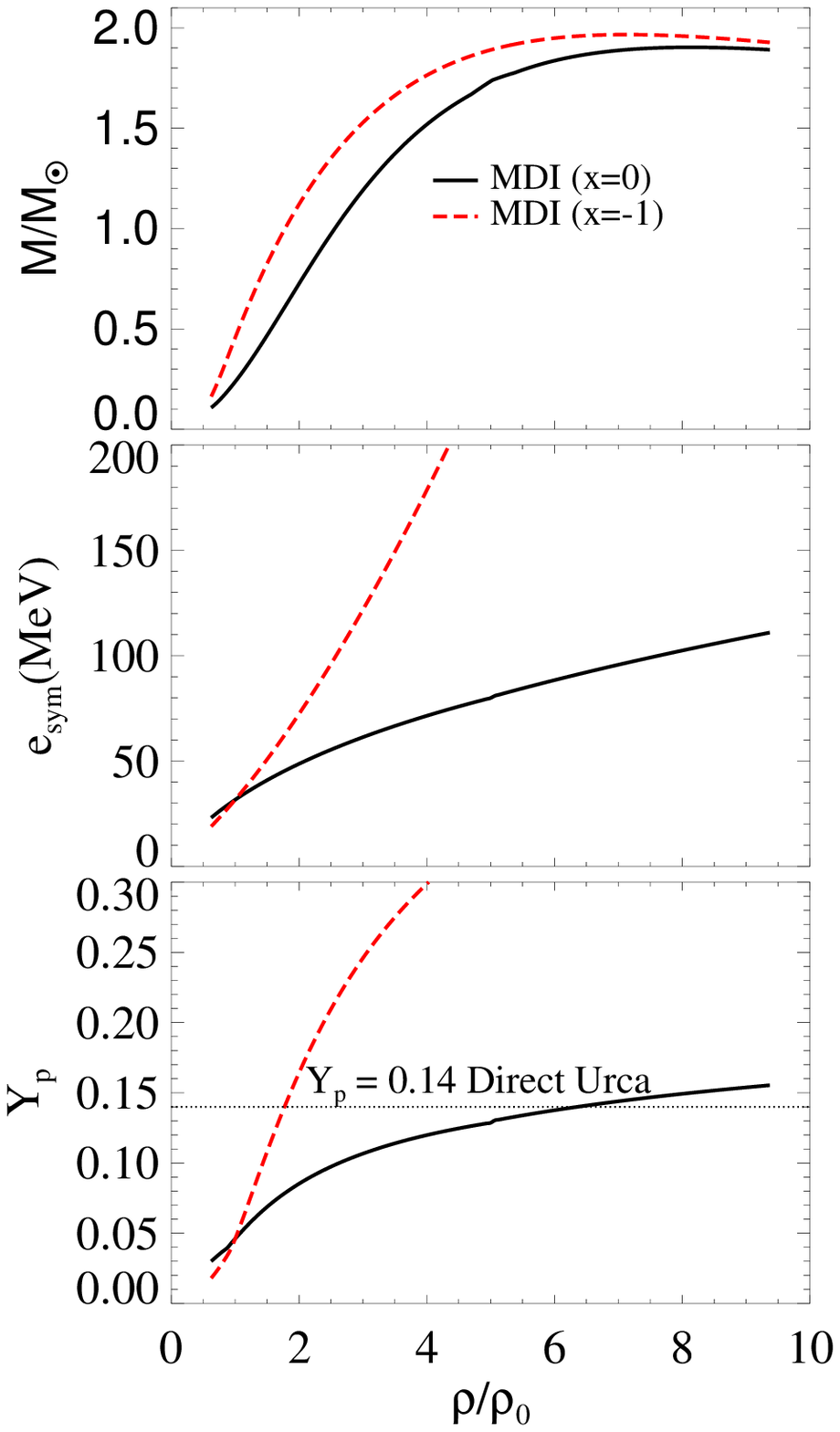}
\vspace{5mm} \caption{(Color online) Neutron star mass (upper
panel), nuclear symmetry energy (middle panel) and proton fraction
(lower panel) versus baryon density. Predictions from the MDI
(x=-1,0) are shown.}
\end{figure}

Finally, we study the effect of (fast) rotation on the proton
fraction in the star core. In Fig.~8 we show the central baryon
density (upper frame) and central proton fraction (lower frame) as a
function of the rotational frequency for fixed-mass models.
Predictions from both $x=0$ and $x=-1$ EOSs are shown. We observe
that central density decreases with increasing frequency. This
reduction is more pronounced in heavier neutron stars. Most
importantly, we also observe decrease in the proton fraction $Y_p^c$
in the star's core. We recall that large proton fraction (above
$\sim 0.14$ for $npe\mu$-stars) leads to fast cooling of neutron
stars through direct Urca reactions. Our results demonstrate that
depending on the stellar mass and rotational frequency, the central
proton fraction could, in principle, drop below the threshold for
the direct {\it nucleonic} Urca channel and thus making the fast
cooling in rotating neutron stars impossible. The masses of the
models shown in Fig.~8 are chosen so that the proton fraction in
stellar core is just above the direct Urca limit for the {\it
static} configurations, see Fig.~9 upper and lower frames. The
stellar sequences in Fig.~8 are terminated at the Kepler
(mass-shedding) frequency. In both cases the central proton fraction
drops below the direct Urca limit at frequencies lower than that of
PSR J1748-244ad \citep{Hessels:2006ze}. This implies that the fast
cooling can be effectively blocked in millisecond pulsars depending
on the exact mass and spin rate. It might also explain why heavy
neutron stars (could) exhibit slow instead of fast cooling. For
instance, with the $x=0$ EOS (with softer symmetry energy) for a
neutron star of mass approximately $1.9M_{\odot}$, the Direct Urca
channel closes at $\nu\approx 470Hz$. On the other hand, with the
x=-1 EOS (with stiffer symmetry energy) the direct Urca channel can
close only for low mass neutron stars, in fact only for masses well
below the canonical mass of $1.4M_{\odot}$. This is due to the much
stiffer symmetry energy (see Fig.~9 middle frame) because of which
the direct Urca threshold (Fig.~9 lower frame) is reached at much
lower densities and stellar masses (Fig.~9 upper frame).

Before closing the discussion in this section a few comments are in
order. In the present study we do not consider "exotic" states of
matter in neutron stars. On the other hand, due to the rapid rise of
the baryon chemical potentials several other species of particles,
such as strange hyperons $\Lambda^0$ and $\Sigma^-$, are expected to
appear once their mass thresholds are
reached~\citep{2000PhRvC..61e5801B}. The appearance of hyperonic
degrees of freedom lowers the energy-per-particle in the stellar
medium and causes more centrally condensed configurations with lower
masses and radii. Additionally, hyperons help the condition for {\it
nucleonic} direct Urca process to be satisfied at lower densities
due to the increased proton fraction \citep{1992ApJ...390L..77P},
and depending on their exact concentrations could potentially
contribute to the fast cooling of the star through {\it hyperonic}
direct Urca processes \citep{Page:2005fq}. These considerations
would alter the balance between the curves in Fig.~9 and ultimately
the results displaced in Fig.~8 in favor of the direct Urca process,
i.e. smaller masses and higher frequencies would be necessary to
close the fast cooling channel. This is due to the fact that the
overall impact of (rapid) rotation on the neutron star structure is
smaller for more centrally condensed models resulting from
``softer'' EOSs \citep{1984Natur.312..255F}. Therefore, for such
models there is smaller deviation from properties and structure of
static configurations. In addition, at even higher densities matter
is expected to undergo a transition to quark-gluon plasma
\citep{Weber:1999a,2000PhRvC..61e5801B}, which favors a fast cooling
through enhanced nucleonic direct Urca and quark direct Urca
processes \citep[see e.g.,][]{Page:2005fq}.

\section{Summary}

We have studied properties of (rapidly) rotating neutron stars
employing four nucleonic EOSs. Rapid rotation affects the neutron
star structure significantly. It increases the maximum possible mass
up to $\sim 17\%$ and increases/decreases the equatorial/polar
radius by several kilometers. Our findings, through the application
of EOSs with constrained symmetry energy by recent nuclear
terrestrial laboratory data, allowed us to constrain the mass of the
neutron star in XTE J1739-285 to be between 1.7 and $2.1M_{\sun}$.

Additionally, rotation reduces central density and proton fraction
in the neutron star core, and depending on the exact stellar mass
and rotational frequency could effectively close the fast cooling
channel in millisecond pulsars. This circumstance may have important
consequences for both the interpretation of cooling data and the
thermal evolution modeling.

\section*{Acknowledgements}We would like to thank Nikolaos Stergioulas
for making the RNS code available. We also thank Wei-Zhou Jiang for
helpful discussions. This work was supported by the National Science
Foundation under Grant No. PHY0652548 and the Research Corporation
under Award No. 7123.

\end{document}